\documentstyle[12pt]{article}
\global\arraycolsep=2pt

\begin{document}
\begin{titlepage}
\begin{flushright}
SMU-HEP-94-12\\
August  1994
hep-ph/9409251
\end{flushright}

\vspace{0.3cm}
\begin{center}
{\Large\bf Axion search with optical technique.}
\end{center}
\vskip 0.3cm

 \begin{center} {\bf Ryszard Stroynowski\footnote{E-mail:
ryszard@mail.physics.smu.edu.}
and Ariel R. Zhitnitsky\footnote{
On leave of absence from Budker Institute of Nuclear Physics,
Novosibirsk, 630090 Russia\\E-mail: arz@smuphy.physics.smu.edu.}}
 \end{center}
\begin{center}
{\it Southern Methodist University, Dallas, Texas 75275, USA.}
\end{center}
\vskip 2cm
\begin{abstract}
A brief review of particle physics motivations for axion existence is
followed by a discussion of optical observables that can be used for 
axion detection. It is emphasised that the resonant photon - axion transition
can occur in the presence of low density 
gas bringing the experimentally accessible
quantities within the reach of the present day technology. 

\end{abstract}
\end{titlepage} 
\vskip 1cm
{\bf 1. Introduction.}
\vskip 0.1in
 Recently, there has been a renewed interest \cite{prop} 
in the possibility of searching for 
light pseudoscalar particles using laser interferometry. This interest has
been stimulated by the availability of SSC and other 
magnets for non accelerator
applications. The idea of using of the optical techniques to measure 
vacuum birefringence
in the presence of strong magnetic fields and to search for scalar and
pseudoscalar particles was originally proposed by Iacopini and Zavattini 
\cite{iacopz} and has been extensively reviewed and discussed by several authors
mainly in the context of the search for solar axions \cite{Sikivie},
\cite{Maiani},\cite{Bibber},\cite{Lazarus},\cite{Stodolsky}.
A discussion of the QED tests can be found in
\cite{Ni}. 
First application of this technique was made by an experiment of 
Cameron et al., \cite{Cameron} which used laser
interferometry to measure optical rotation of light polarization
in a magnetic field and was able to estimate a limit for the axion coupling 
to two photons 
$ g_{a\gamma\gamma}< 3.6\cdot 10^{-7}$ Ge$V^{-1}$ for axion masses
$m_a < 10^{-3}$ eV.       

In this note we first briefly discuss the particle physics motivations for 
the existence of axions and then review the experimental observables and their
limitations for axion detection with optical techniques.
 A complete axion - photon mixing formula is derived without any
approximations. A resonant photon-axion
transition in the presence of low density gas enhances the values of
measurable quantities and suggests a novel experimental approach.
It is shown that, in the resonant case, present technology can reach 
theoretically expected limits
of the photon-axion coupling constant for axion masses $m_a\geq 10^{-3}$ eV.
Astrophysical arguments constraining axion properties have been extensively
reviewed elsewhere \cite{rev}.
\vskip 0.2in
{\bf 2. Review of axion physics.}
\vskip 0.1in
{\bf 2.1 The strong CP problem}
\vskip 0.1in
The axion is generally thought to be the most attractive 
 solution to the strong CP problem. 
We begin by first formulating the problem.

The general Lagrangian for a renormalizable, gauge invariant quantum field
theory that includes only fundamental gauge field $A_{\mu}$ with spin $1$
(photons, gluons...) and fields of matter $ q^i$ 
with spin $1/2$ can be written as:
\begin{equation}
\label{qcd}
L=\frac{-1}{4}F_{\mu\nu}F^{\mu\nu}+
\sum_{i=flavor}\bar{q^i}(i\gamma_{\mu}D_{\mu}-m^i)q^i+
\frac{g^2\theta}{32\pi^{2}}\tilde{F}^{\mu\nu}F_{\mu\nu}
\end{equation} 
Here, $$D_{\mu}=\partial_\mu - igA_\mu$$ is a covariant derivative,
$g$ is the coupling constant,
$$  F_{\mu\nu}\equiv \frac{-1}{ig}[D_{\mu},D_{\nu}] $$ 
is the strength tensor of the gauge field and 
$$\tilde{F}_{\mu\nu}=\frac{1}{2}\epsilon_{\mu\nu\lambda\sigma}
F^{\lambda\sigma}$$ is the dual of the field strength
tensor.
 The third term in the Lagrangian,
characterized by the strength $\theta$, is a total derivative; it 
  affects  neither the equations of motion, nor the perturbative 
aspects of the theory.

In the  case of Quantum Electrodynamics 
the coupling constant $g$ should be identified
with electric charge and $A_{\mu}$
 with photon field.
Because QED
  is a purely perturbative theory,
the theta term  can be disregarded since it does not contribute to any
observable quantity.
 
In the Standard Model of electroweak interactions,
some nonperturbative effects might occur. However, their
contributions, characterized by the $\theta$ term, are proportional to     
$  e^{-\frac{2\pi}{\alpha}}$, where the fine structure constant 
$\alpha\equiv\frac{g^2}{4\pi}=
\frac{1}{137}$ .
This proportionality factor is a very small number 
and can be neglected in general
discussions. 

In the case of strong interactions, 
the coupling   $g$ in the formula (\ref{qcd}) is related to a
strong coupling constant $\alpha_s$ via $\alpha_s=\frac{g^2}{4\pi}$; the field 
$A_{\mu} $ should be identified 
 with an octet of gluon fields  and $ F_{\mu\nu}$
is the 
strength tensor of the gluon field.
Because of a large 
value of the strong coupling constant 
$\alpha_s $, this term should have, in principle, a 
significant
contribution. In fact, in QCD it can be proven 
that the observable quantities depend on $\theta$.
 
 The existence of the $\theta$ term implies a violation of P,
CP and T symmetries.
However,  
there is no experimental 
evidence for  P or CP violation in strong interactions. For 
example, CP violation in QCD would induce electric dipole moments of
strongly interacting particles and there are stringent experimental
limits on those quantities. Thus the absence of CP violating effects in QCD
indicates a very small value for the parameter $\theta$. The most restrictive 
upper limit on its value can be
derived from the experimental bound   \cite{PDG}
on the neutron electric dipole moment:  
$d < 12\cdot 10^{-26}$ e-cm at 95 \% CL.  
 The corresponding upper limit on $\theta$ is \cite{Witten}:
\begin{equation}
\label{theta}
\theta\leq 10^{-9}.
 \end{equation} 

Now we are ready to formulate the  problem: why is $\theta$ so small?
 
There are several possible answers to this question.
The most elegant one was proposed by Peccei and Quinn who assumed that
the strong interactions 
Lagrangian has a global $U_{PQ}(1)$ chiral symmetry \cite{PQ}.
 Weinberg and Wilczek   \cite{WW}
 analysed the consequences of the Peccei-Quinn symmetry
and  noticed that the  spontaneous breaking 
of a global chiral symmetry $U_{PQ}(1)$
  leads to a light pseudoscalar
pseudo-Goldstone boson, called an axion, that will interact with topological
charge.             
 
To see, how the existence of a $U_{PQ}(1)$ symmetry automatically leads
to the effective $\theta_{eff}=0$ let us 
introduce into  theory a new scalar field $\phi(x)$
in such a way that, under a $U_{PQ}(1)$ transformation, 
the quark and $\phi$ fields
  transform as:
\begin{equation}
\label{pq}
 \phi\rightarrow  e^{i\alpha} \phi , ~~~ q_i\rightarrow 
e^{-i\alpha\gamma_5 /2} q_i.
\end{equation}
In other words, we allow for existence of an additional complex scalar
field,
which must be included in the Lagrangian. In this case,
the phase of the new field  cancels exactly
the original $\theta$ parameter.
The price for this cancellation
is an appearance of a new particle - the axion {\it a}.

The axion couples
to all constituents of matter - quarks and gluons. The coupling strength
to fermions is proportional to $m_q\over{f_a}$, where $m_q$ represents
a quark mass and $f_a$  is a dimensional parameter
describing the energy scale of the  
Peccei-Quinn symmetry
breaking.
All physical observables can be expressed in terms of $f_a$.
% but also to gluons.
For example, the axion's interactions with gluons 
are described by a triangle diagram, shown in Fig.1
%\vskip 2.0in
%\centerline{Fig.1}
and are given by 
\begin{equation}
\label{axlna}
L_{agg}=
\frac{g^2}{32\pi^{2}}\tilde{F}^{{\mu\nu}a}
F_{\mu\nu}^a\bigl(\frac{a}{f_a}\bigr).
\end{equation}
 These interactions 
mix the
axion with the neutral pion through the gluons, 
yielding an effective low-energy axion potential.
%This potential forces the axion field into its
%CP-conserving minimum $\theta_{eff}=\theta-\alpha=0$ and
It generates an axion  mass in analogy to the way in which the chiral symmetry
breaking generates a mass for the $\pi$ meson.

\vskip 0.1in 
{\bf 2.2 Axion mass}
\vskip 0.1in
The axion mass is related to the QCD scale parameter via
   $$m_a\sim\Lambda^2_{QCD}/f_a,$$
where $\Lambda_{QCD}$ represents a scale of strong interactions physics
and is of the order of 100 MeV.

Originally, $f_a$ was thought to be the same as the 
symmetry-breaking scale of the weak interactions, i.e., about 250 GeV.
As a consequence the axion mass was expected be of the order of 100 keV. 
This type of axion has been ruled out 
early by particle decay and beam dump experiments.
However, allowing the Peccei-Quinn symmetry breaking scale $f_a$ 
to become much 
larger results
in an axion with much lower mass and a weaker coupling to matter.
Models for such "invisible axions" include
the Dine-Fischler-Srednicki-Zhitnitsky (DFSZ) axion \cite{DFSZ}
which couples to all fermions and the 
Kim-Shifman-Vainshtein-Zakharov (KSVZ) axion \cite{KSVZ}
which couples  to quarks only.

For invisible axions there are no a priori bounds
on $f_a$. However, 
%because the axion is associated with the
%spontaneous breaking of the Peccei-Quinn symmetry, 
one can calculate
the relevant properties of this particle in terms of $f_a$.
The axion mass
comes from the mixing of the axion with the neutral pion.
Thus the axion mass is proportional
to $1/f_a$ and to the neutral pion parameters: its mass
$m_{\pi}=135$ MeV and its decay constant $f_{\pi}=133$ MeV.
 The exact relation is given by \cite{rev}:
  \begin{equation}
\label{mass}
m_a=\frac{f_{\pi}m_{\pi}}{f_a}\frac{\sqrt{m_um_d}}{(m_u+m_d)}
\simeq 0.6\cdot 10^{-5}eV\bigl(\frac{10^{12} GeV}{f_a}\bigr)
 \end{equation} 
where $m_u$ and $m_d$ represent masses of up and down quarks making up the
neutral pion.
For the numerical estimate we assume $m_u=4$ MeV and $m_d=7$ MeV.
%as the standard parameters for these light quarks.
\vskip 0.1in
{\bf 2.3 Axion interactions}
\vskip 0.1in
The most important quantity for us is the coupling 
  of the axion to two photons,
described by the coupling constant $g_{a\gamma\gamma}$, which
can be derived from the axion-fermion interactions.
There are two independent contributions to 
$g_{a\gamma\gamma}$. One of them comes from interactions
with leptons (electrons, muons and taus)
through a triangle loop. Because 
the axion coupling with fermions is proportional to
$1/f_a$ and is very small, $g_{a\gamma\gamma}$ is small. 
This contribution
is exactly zero for the KSVZ model, in which the axion-lepton
  interactions are absent.
The second - a hadronic contribution - is related to the 
mixing of axion with $\pi^0$ meson. 
%As we mentioned above this mixing
%is very small $\sim 1/f_a$, thus, the corresponding
%contribution to $g_{a\gamma\gamma}$ will be suppressed by
%the same factor. 
The exact formula \cite{rev} for $g_{a\gamma\gamma}$
 is given by:
\begin{equation}
\label{photon}
g_{a\gamma\gamma}=
\frac{\alpha}{2\pi f_a}\cdot(\frac{N_l}{N}-
\frac{5}{3}-\frac{m_d-m_u}{m_d+m_u}),
  \end{equation} 
where $g_{a\gamma\gamma}$ is defined by the following Lagrangian of photon
- axion interactions:
\begin{equation}
\label{photoax}
L_{a\gamma\gamma}=\frac{g_{a\gamma\gamma}}{4}\tilde{F}_{\mu\nu}
F_{\mu\nu}\cdot a=-\frac{1}{M}\vec{E}\cdot\vec{B}a.
\end{equation}
Here, we see explicit relation of the Lagrangian to 
the electric field $\vec{E}$ and the
magnetic field $\vec{B}$. The axion interacts only with the 
photon wave component
 parallel to an external magnetic field.
 The energy scale $M$ is defined as
$$ M \equiv\frac{1}{g_{a\gamma\gamma}}.$$
     
       In the above definitions, $N_l$ is a constant 
%related to electromagnetic 
%triange loop diagram and 
proportional
to the number of charged leptons and the factor $N$ is an analogous constant
related to the number of gluons. For each model
these constants can be explicitly calculated. The 
  ratio $\frac{N_l}{N}$   differs from model to model.
%The second contribution to $g_{a\gamma\gamma}$,
%as we discussed, is related to axion-pion mixing
%and is one and the same for both kind of axions. 
In the DFSZ model, as well as in all grand unified
axion models,
% which obtain the successful prediction
%for Weinberg angle  $\sin^2\theta_W$, one has 
the value of the ratio $\frac{N_l}{N}$ is \cite{rev}: 
$$\frac{N_l}{N}=\frac{8}{3}.$$ 
 In the  (KSVZ) model there are no axion interactions with
leptons ($N_l=0$).
Therefore, the relative strength of the axion-photon-photon coupling
in the two models is given by $\xi$:
\begin{equation}
\label{xi}
\xi\equiv\frac{g_{a\gamma\gamma}(DFSZ)}{g_{a\gamma\gamma}(KSVZ)}
=\frac{\frac{8}{3}-\frac{5}{3}-\frac{m_d-m_u}{m_d+m_u}}
{-\frac{5}{3}-\frac{m_d-m_u}{m_d+m_u}}\simeq -0.37.
\end{equation} 
Bearing in mind that all observable values depend on 
$g_{a\gamma\gamma}^2$ we conclude that KSVZ-axion
interactions are  $\xi^{-2}\simeq 7$ times stronger 
than DFSZ-axion interactions.

Equations (\ref{mass}) and (\ref{photon}) also show that the axion
mass $m_a$  and axion-photon coupling constant $M^{-1}$
 are not independent parameters,
but are related to each other by:
\begin{eqnarray}
\label{mM_dfsz}
m_aM (DFSZ) =7.4\cdot 10^{18}eV^2, 
\rightarrow (\frac{m_a eV}{10^{-5}})
(\frac{M GeV}{10^{15}})=0.74   
\end{eqnarray}
\begin{eqnarray}
\label{mM_ksvz}
m_aM (KSVZ) =
 2.7\cdot 10^{18}eV^2
\rightarrow (\frac{m_a eV}{10^{-5}})(\frac{M GeV}{10^{15}})=0.27
\end{eqnarray} 

A priori the axion mass window spans the range $10^{-12}$ eV to $10^6$ eV.
Laboratory experiments searched without success 
for axions with mass $m_a>10$ keV in the decays
$K^+\rightarrow\pi^+ + a$, ~$J/\psi\rightarrow a + \gamma$,
~$\Upsilon\rightarrow a + \gamma$ and others.
Both high-mass and low-mass axions are constrained by  
astrophysical arguments originally advanced by Dicus {\it et al.,}
\cite{Dicus}
and extensively discussed in excellent reviews \cite{rev}. 
Present understanding of the evolution of red giants eliminates
the window of $10^{-2}$eV$<m_a<10^5$eV while the observed duration
of the neutrino burst from the SN1987A supernova eliminates
$10^{-3}$eV$<m_a<2$ eV. Very low mass axions with $m_a<10^{-6}$ eV
are excluded by cosmological considerations. The main
window of axion mass still allowed  is in the range
$10^{-6}eV\leq m_a\leq 10^{-3}eV$.
\vskip 0.1in 
{\bf 2.4 Axion lifetime}
\vskip 0.1in
The axion width for decay into two photons is defined by the 
axion-photon interaction
(\ref{photoax}) and is given by
\begin{eqnarray}
\label{time}
 \Gamma(a\rightarrow\gamma\gamma)=\frac{g_{a\gamma\gamma}^2m_a^3}{64\pi}.
\end{eqnarray}
Its lifetime depends on its mass (or on the symmetry breaking scale M):
\begin{eqnarray}
\tau(a\rightarrow\gamma\gamma)\sim 5\cdot10^{49}(\frac{M ~GeV}{10^{15} ~GeV})s, 
\end{eqnarray}
\noindent
which is sufficiently long to have no effect on our observations.
\vskip 0.1in 
{\bf 3. Definitions of observables.}
\vskip 0.1in

In order to define the relations between the physical processes involved in
light propagation in a magnetic field and the observable quantities, let us
consider a light wave, with an amplitude $\vec{A}$ and angular frequency
$\omega$, that initially is linearly polarized
at an angle $\phi$ with respect to an external magnetic field $\vec{B}$:
\begin{eqnarray}
\label{1}
\vec{A}=\vec{i}\cos\phi+\vec{j}\sin\phi,
\end{eqnarray}
where $\vec{i}$ is unit vector parallel to $\vec{B_{ext}}$,
 $\vec{j}$ is orthogonal to $\vec{B_{ext}}$ and $\vec{k}$ is the direction of
propagation.
At a time t the amplitude can be expressed as:
\begin{eqnarray}
\label{2}
\vec{A}=\bigl(\vec{i}A_{\parallel}(t)\cos\phi+\vec{j}A_\perp(t)\sin\phi\bigr)
e^{-i(\omega t-\omega\vec{k}z)}.
\end{eqnarray}
The propagation of light in the magnetic field is affected by several different
processes. The most significant is a pure, second order in $\alpha$, 
QED process of Delbruck scattering
off the
virtual photons coupled to virtual electron loops shown in Fig.2.
%\vskip 2in
%\centerline{Fig. 2}
This is a process in which the vacuum can be treated as a sea of virtual
electron-positron pairs. In an external magnetic field these pairs become 
polarized and the resulting polarized electric field interferes with the
electric field of a propagating photon. The interference affects
the parallel and perpendicular components differently
and results
in a change of the polarization of the light wave.
This effect is referred to as the
QED vacuum birefringence and it is derived from
the  Heisenberg effective Lagrangian \cite{Heis}:
\begin{equation}
\label{Heis}
L=\frac{-1}{4}F_{\mu\nu}F_{\mu\nu}+
 \frac{\alpha^2}{90m_e^4}[(F_{\mu\nu}F_{\mu\nu})^2+
\frac{7}{4}(\tilde{F_{\mu\nu}}F_{\mu\nu})^2]
\end{equation} 
The same Lagrangian can be expressed in terms of explicit electric and magnetic
fields:
\begin{equation}
\label{heisfield}
L=\frac{1}{2}({\vec{E}}^2-{\vec{B}}^2)+\frac{2\alpha^2}{45m^4_e}
\bigl[{\vec{E}}^4+{\vec{B}}^4-2{\vec{E}}^2{\vec{B}}^2+7(\vec{E}\cdot
\vec{B})^2\bigr]
\end{equation}
Another process which may influence the light propagation is related to
photon - axion interactions.
Equation (\ref{photoax}), describing these interactions, contains
a dot product of the electric and magnetic fields. Therefore, a photon with its
electric field  $\vec{E}$ polarized in the direction parallel to an 
external magnetic
field  $\vec{B_{ext}}$ can produce axions. A photon with $\vec{E}$ polarized
perpendicular to the magnetic field remains unaffected.
For example, if a light beam enters
the magnetic field region with its linear polarization at an angle with respect 
to the magnetic field direction, its parallel component $A_{\parallel}$
is attenuated and undergoes a change of phase (because of axion production), 
while its 
orthogonal component remains the same. 

Neglecting effects that are even higher order in $\alpha$, we can
 use an approximate expression for the parallel component of the 
amplitude, $A_{\parallel}(t)$, that separates the phase shift of the wave from
its attenuation:
\begin{eqnarray}
\label{3}
  A_{\parallel}(t)=(1+\epsilon_{\parallel}(t)+i\varphi_{\parallel}(t) ).
\end{eqnarray}
\begin{eqnarray}
\label{456}
  A_{\perp}(t)=(1+i\varphi_\perp(t))
\end{eqnarray}
In this representation $\epsilon(t)$ describes the attenuation and 
$\varphi_{\parallel}(t)$ and $\varphi_\perp(t)$ are related to the phase shift.
The physical quantities that can be measured
are:
\vskip 0.1in
{\bf ellipticity:} $\psi=\frac{1}{2}|\varphi_{\parallel}(t)-\varphi_{\perp}(t)|
\sin2\phi$ 
\vskip 0.1in
{\bf rotation:} $\varepsilon=\frac{1}{2}|\epsilon_{\parallel}(t)|\sin2\phi$.
\vskip 0.1in
The ellipticity describes the ratio of the minor and major axes of the photon
polarization vector. It 
depends on the 
phase shift of the light wave and is generated by both
the pure QED and photon-axion interference. 
The rotation refers to a rotation of the photon polarization vector after the 
passage through the magnetic field. It depends on the
wave attenuation only and cannot be generated by pure QED processes.
Polarized light propagating over a path length $l$ through a transverse
magnetic field will gain an ellipticity due
to Delbruck scattering given by: 
\begin{equation}
\label{psi}
\psi_{QED}=\frac{\alpha^2 B_{ext}^2l\omega}{15m_e^4}.
 \end{equation} 
\begin{equation}
\label{epsqu}
\varepsilon_{QED}=0 .
\end{equation}
If the light passes through the same magnetic field N times, the corresponding
formula for ellipticity is:
\begin{equation}
\label{psi1}
\psi_{QED}=\frac{\alpha^2 B_{ext}^2l\omega{N}}{15m_e^4}.
 \end{equation} 
Here, and throughout the paper, we use a unit system in which $\hbar=c=1$. 
In this system
the length and magnetic field units can be expressed in terms of eV:
$$1 m = 5\cdot 10^6 ~eV^{-1}$$
$$1 T = 195 ~eV^2$$

It is often convenient to present these results
in terms of refractive indices $n_{\perp}$ and $n_{\parallel}$
which refer to the transverse and parallel polarization of the wave with respect
to an external magnetic field.
\begin{equation}
\label{n}
   ~~~~~n_{\perp}=\frac{c}{velocity~ of~ transverse~ wave}\equiv 
1+\frac{\Delta_{\perp}}{\omega}
\end{equation}
and
\begin{equation}
\label{n1}
  ~~~~~n_{\parallel}=\frac{c}{velocity~ of~ parallel~ wave}\equiv 
1+\frac{\Delta_{\parallel}}{\omega} .
\end{equation}
Here, the index of refraction is defined \cite{Jackson} by the relation between
 the photon momentum $k$ and its frequency $\omega$
\begin{equation}
\label{xyz}
k=n\cdot\omega ,
\end{equation}
where the velocity of light in vacuum is $c = 1$.
In terms of $\Delta_{\parallel}$ and $\Delta_{\perp}$ the ellipticity
is given by $\psi=\frac{1}{2}\frac{|\Delta_{\parallel} -\Delta_{\perp}|}
{\omega}\sin2\phi$.

In practice, a perfect vacuum does not exist. Therefore
the $\Delta_{\parallel}$ and $\Delta_{\perp}$ terms in the above equations 
should be extended to represent total refractive indices that include both
vacuum and residual gas contributions.
We shall use the notation $\Delta^{vac},~\Delta^{gas}$, introduced
in ref.\cite{Stodolsky}, for the vacuum and gas
contributions respectively.
$\Delta^{gas}$ is a  function of density, temperature,
chemical composition of the gas, and magnetic field strength.
In an external magnetic field
$\Delta_{\parallel}^{gas} \neq\Delta_{\perp}^{gas}$, and
the difference contributes to the so called Cotton-Mouton effect of
birefringence in gasses and liquids in a magnetic field.
The effect can be accounted for by a Cotton-Mouton constant C defined by
\begin{equation}
\label{CM}
 \frac{\Delta_{\parallel}^{gas} -\Delta_{\perp}^{gas}}{\omega}=
 n_{\parallel}^{gas} -n_{\perp}^{gas} =C\lambda B_{ext}^2,
\end{equation}
where $\lambda$ is the wavelength of
the light.
  \vskip 0.1in
{\bf 4. Axion-photon mixing.}
\vskip 0.1in
We are ready now to formulate the result for photon-axion mixing
in an external field $B$. 
The axion mixes only with a photon component that is parallel to an
external magnetic field. The physical states resulting from the mixing
are linear combinations of the axion and the 
parallel component of the photon wave:
%In order to do so, let us
%introduce the   mixing angle $\vartheta$ in the following way:
$$A_{\parallel}'=A_{\parallel}\cos\vartheta+a\sin\vartheta$$
$$a'=-A_{\parallel}\sin\vartheta+a\cos\vartheta .$$
The strength of the mixing is characterized  by a mixing angle $\vartheta$ 
% by  few essential  physical parameters  
\cite{Stodolsky}:
\begin{equation}
\label{3.1}
\frac{1}{2}\tan(2\vartheta)=\frac{B }{2M(\Delta_{\parallel}-\Delta_a)}.
\end{equation}
The parameter $\Delta_a$ is related to the axion mass and photon frequency by
$$\Delta_a = -\frac{m_a^2}{2\omega}. $$
The sign of the $\Delta_a$ comes from the dispersion relation
$$k=\sqrt{\omega^2-m_a^2}\simeq\omega (1-\frac{m_a^2}{2\omega^2}).$$
Finally, we can introduce a concept of oscillation length described by:
 $\frac{2\pi}{\Delta_{osc}}$, where
$\Delta_{osc}=\Delta_{\parallel}-\Delta_a$.
The mixing angle $\vartheta$ in most cases is very small.

We will now discuss the effects on the light observables - ellipticity
and rotation - from the axion-photon mixing. 
We discuss the case in which light propagates in a magnetic field
of length l. In some experiments, light bounces N times through 
the length l and the corresponding effects increase linearly with N.
\vskip 0.1in
 {\bf 4.1 Axion-photon mixing in vacuum.}
\vskip 0.1in

We follow calculations presented by Raffelt and Stodolsky \cite{Stodolsky}.
The axion contribution to ellipticity is:
\begin{eqnarray}
\label{6}
 \psi_a=\frac{1}{2}\frac{{\omega}^2 B_{ext}^2 N}{M^2m_a^4}(\gamma-\sin\gamma).
\end{eqnarray}
Here, $\gamma$ is a dimensionless parameter relating path length, mass of the
axion and the angular frequency of the photon:
$$\gamma\equiv\frac{m_a^2l}{2\omega}.$$ 

The analogous formula for 
  rotation  is:
\begin{eqnarray}
\label{7}
 \varepsilon_a= \frac{{\omega}^2 B_{ext}^2 N}{M^2m_a^4}\cdot 
\sin^2(\frac{\gamma}{2}),
\end{eqnarray}
where $\gamma$ is given above.
For  $\gamma\ll 1$ , which occurs for small axion masses and
relatively short magnetic field length,
one can expand
the above formulae (\ref{6}, \ref{7})  to get  
simple approximate expressions:
\begin{eqnarray}
\label{8}
 \psi_a= \frac{m_a^2 B_{ext}^2l^3 N}{96\omega M^2 }, ~~~\gamma\ll 1 
 \end{eqnarray}
\begin{eqnarray}
\label{9}
 \varepsilon_a= \frac{l^2 B_{ext}^2 N}{16M^2 }, ~~~~~~\gamma\ll 1.
\end{eqnarray}
\vskip 0.1in
{\bf 4.2 Axion-photon mixing in media.}
\vskip 0.1in
Up to now we discussed  axion-photon mixing in vacuum, where the
effect is weak. The mixing is more complicated
for  photon propagation through a medium. In that case we have to use 
a complete formula for the mixing.
General expressions for the axion-photon mixing contributions to the
ellipticity and rotation can be derived following the same method as used 
by Raffelt and Stodolsky \cite{Stodolsky} but without the small mixing angle 
approximation. The complete solution is obtained by diagonalization of the
mixing matrix:
\begin{equation}
\label{ful1}
\varepsilon=\varepsilon_{QED}+\frac{1}{2}\bigl[1-\cos^2\vartheta\cos\bigl(\frac{Bl}{2M}
\tan\vartheta\bigr)-\sin^2\vartheta\cos\bigl(\frac{Bl}{2M}\cot\vartheta
\bigr)\bigr]
\end{equation}
\begin{equation}
\label{ful2}
\psi=\psi_{QED}+\frac{1}{2}\bigl[\cos^2\vartheta
\sin(\frac{Bl}{2M}\tan\vartheta\bigr)-\sin^2\vartheta\sin(\frac{Bl}{2M}
\cot\vartheta\bigr)\bigr]
\end{equation}
In what follows, we explore the regions of  mixing angle in
which mixing effects are largest.

% make no assumptions on the 
%mixing parameter.
For small values of the mixing angle, $ \vartheta\ll 1 $, we can use the 
approximation 
$$\frac{1}{2}\tan(2\vartheta)\simeq \vartheta$$
and the formulae for the ellipticity and rotation
in terms of $\vartheta$ are analogous to
previous discussed equations (\ref{6}) and (\ref{7}):
\begin{equation}
\label{3.2}
\psi_a= \frac{\vartheta^2N}{2}[\Delta_{osc}l-\sin(\Delta_{osc}l)]
\end{equation}
\begin{equation}
\label{3.3}
\varepsilon_a=  \vartheta^2N \sin^2(\Delta_{osc}l/2)
\end{equation}
where
$$\vartheta=\frac{B }{2M\Delta_{osc}}\ll 1$$
and
$$\Delta_{osc}=\Delta^{vac}_{\parallel}+\Delta^{gas}_{\parallel}
-\Delta_a\simeq\frac{m_a^2}{2\omega}$$

\vskip 0.2in
{\bf 4.3 Resonant case.}
\vskip 0.1 in 

Another region of interest is the axion-photon resonance,
 which occurs when $\Delta_{\parallel}\simeq\Delta_a $.
In vacuum the signs of the pure QED and axion contributions are opposite
$\Delta_{\parallel}^{vac}>0$ and $\Delta_a <0$ and their difference is always
positive. However, in a gas medium with a plasma frequency much smaller 
then the photon frequency, the light scattering contribution is negative 
$\Delta_{\parallel}^{gas}< 0$. In such case 
it is possible to obtain a cancellation in the denominator of the
Eq. (\ref{3.1}) when
$\Delta_{\parallel}-\Delta_a =\Delta^{vac}_{\parallel}+\Delta^{gas}_{\parallel}
-\Delta_a=0$. This corresponds to a resonant condition $\vartheta=45^0$.
 In the ideal resonance case 
the expression for the rotation is given by
\begin{equation}
\label{3.4}
\varepsilon_a=  \frac{1}{2}(1-\cos(\frac{Bl}{2M}))\simeq
\frac{1}{4}(\frac{Bl}{2M})^2
\end{equation}
The photon-axion transition probability is given by
\begin{equation}
\label{3.5}
P(\gamma_{\parallel}\rightarrow a)=  \sin^2(\frac{Bl}{2M}).
\end{equation}
All these formulae should be multiplied by a factor $N$
for multiple-beam-path experiments.
>From these formulae it is seen that if we had
the length $l_{ocs}\sim\frac{\pi M}{B}$, complete
transition between photons and axions would occur.

The resonant case provides an important handle on axion detection with
optical interferometry.                                     
Naively, one can expect that an increase
in the light propagation path length $l$ through the magnetic field, 
would increase the values of observable quantities. 
This is true for ellipticity from a pure QED process (see eq.(\ref{psi})).
It is also true for ellipticity from the axion contribution
(see eq.(\ref{3.2})), which is almost ten order of magnitude smaller
than that from QED and thus difficult to observe.
However, the rotation $\varepsilon_a$ is free from the QED corrections. 
The expression derived for the rotation dependence on the length
shows an oscillatory behaviour with increasing value of $l$.
This is illustrated in Fig. 3, where the expected value for the rotation is
plotted as function of the length parameter $l$ for axion mass of $10^{-4}$ eV 
for the case of a possible experiment with no additional medium 
discussed in the following section. As can be seen, even the most optimistic 
values of the rotation are very small and are well below experimentally
accessible range. Thus, the only avenue open to experimentation is the case 
of the resonant transition.

 The resonance condition for the photon-axion mixing requires that
the axion and the photon waves are coherent in the 
the volume with nonzero magnetic field. For axions with mass
$m_a\neq 0$, this coherence condition requires the presence of $\Delta_{gas}$
and can be achieved by filling
the region where magnetic field is present with an appropriate medium.
In such a case the photon acquires an effective mass $m_{\gamma}$ which can 
be made to match $m_a$ giving a resonance condition
$$m_a=m_{\gamma}.$$
 
For short wavelength, i.e., energetic photons,
the dispersion relation is given \cite{Jackson} by
$$k^2={\omega}^2-m_{\gamma}^2,$$
where $m_{\gamma}$ plays the role of the
plasma frequency in the medium:
$$m_{\gamma}^2= \frac{4\pi \alpha N_e}{m_e}.$$ Here, $N_e$  
is  the electron density and $m_e$ is the electron mass. 
In general, the effective photon mass
in a gas can be expressed in terms of atomic scattering amplitudes
and we refer the reader to the corresponding literature \cite{Lazarus}, 
\cite{Henke}
on this subject. 
%Here we would like to formulate  the  precise 
%condition for the resonant case in terms of the 
%medium properties as explained above:
The electron density, $N_e$, is proportional
to the gas pressure, $P$, and inverse temperature,
$T^{-1}$, so we 
can rewrite the resonant condition in terms of the gas pressure.
The normalization has to be provided by a measurement. For example,
for helium
gas at room temperature and at 1 atm pressure $m_\gamma$  was found 
\cite{Lazarus} to be 
$\sim 0.3$ eV. The scaling relation gives 
\begin{equation}
\label{3.6}
 P=1 atm \cdot(\frac{m_a~eV }{0.3 ~eV})^2\cdot(\frac{T}{300K}).
\end{equation}
For an axion with mass $m_a=10^{-4}$eV, 
the helium pressure generating a resonance condition
at temperature $T\simeq 4K$ is $P\simeq 10^{-6}$ Torr (see Fig.4). 
The above formula is applicable only in the
limit $\omega\gg m_{\gamma}$, which is satisfied
for any laser with frequency $\omega$ of order few $eV$.
\footnote{An analogous idea
has been discussed for the solar axion search experiment
\cite{Bibber}. The only difference is the range
of the axion mass,
which in that case was considered to be about few eV. The  required 
gas pressure was in the range of $0.1-300$ atm. 
Because of the high energy of solar axions (about few keV) the condition
$\omega\gg m_{\gamma}$ was also satisfied.}
The difference between resonant and non-resonant conditions is illustrated
in Fig.5, where the expected value of optical rotation plotted as
function of the axion mass for the two cases
and for the same values of external magnetic field, photon
energy and path length. As can be seen, the rotation in the resonant case 
does not show
the oscillatory character and can reach higher  values
for axion mass above $10^{-3}$ eV. This is due to the expected increase of the
photon - axion coupling constant with increase of the axion mass.
As a result, the size of an efect is much enhanced in the resonant case 
bringing the observable quantities within reach of present day technology.
An experiment to search for
axions can be performed by a scan of the gas pressure in the magnetic volume. 

The resonant condition can be maintained so long as the photon mass is close
to the axion mass with the required precision given by the relation
\begin{equation}
\label{3.7}
 \frac{m_a-m_{\gamma}}{m_a}\ll\frac{2\omega}{lm_a^2}.
\end{equation}

This formula depends strongly on the mass. For example,
a search for an axion with $m_a\simeq 10^{-3}$eV
requires adjusting  gas pressure and temperature
with about 10\% accuracy. 
Searching for axions with larger masses requires better
accuracy. 
\vskip 0.2in
{\bf 5. Numerical estimates}
\vskip 0.1in 
In this section we discuss the numerical estimates for the experiment 
recently completed 
by the Brookhaven - Rochester - Fermilab - Trieste group, in  
which the laser light passed up to 250 times through two dipole magnets
providing an average magnetic field value of 2.15 Tesla, and 
compare them to a
proposed experiment using 10 SSC dipoles 
of the Accelerator Sector String Test facility (ASST) with a field 
of 6.76 Tesla and total
active length of 150 meters.
\vskip 0.1in 
{\bf 5.1 Brookhaven-Rochester-Fermilab-Trieste experiment}
\vskip 0.1in
We start with discussion of the Rochester - Brookhaven - Fermilab - Trieste
experiment of Cameron et al.\cite{Cameron}
The relevant parameters are:
\begin{eqnarray}
\label{Cameron}
l=8.8m,~N=250,~B_{ext}^2=4.5T^2,~\omega=2.41eV.
\end{eqnarray}

  For these parameters, the value of $\gamma$,
%indeed small in whole interesting region of $m_a$: 
\begin{eqnarray}
\label{10}
 \gamma=(\frac{m_a}{0.33\cdot 10^{-3} eV})^2 ,
\end{eqnarray}
is small for the mass range $m_a < 10^{-4}$ eV
and we can use the approximate expressions (\ref{8}) and (\ref{9}).
In the absence of systematic errors, the expected values of the ellipticity
and rotation are:
$$\psi_{QED}\simeq 0.24\cdot 10^{-12}$$ 

\begin{eqnarray}
\label{11}
 \psi_a=1.57\cdot 10^{-22}\frac{(\frac{m_a eV}{10^{-3}})^2}
{(\frac{M GeV}{10^{13}})^2},
~~~~\gamma\ll 1. 
 \end{eqnarray}
\begin{eqnarray}
\label{12}
 \varepsilon_a=5\cdot 10^{-23}\frac{1}
{(\frac{M GeV}{10^{13}})^2},
~~~~\gamma\ll 1. 
 \end{eqnarray}
For an axion mass $m_a=10^{-4}$ eV which corresponds to a symmetry 
breaking scale 
M=$0.74\cdot10^{14}$ GeV in DFSZ model, we expect $\psi_a=3\cdot 10^{-26}$ 
and 
$\varepsilon_a=9\cdot{10^{-25}}$. These values are illustrated in Fig.6. They
are exceedingly small, well
beyond the reach of the experiment.

 \vskip 0.1in
{\bf 5.2 ASST experiment}
\vskip 0.1in
In the initial stage of the ASST experiment we expect to have
10 SSC dipole magnets operating at nominal field value and a
long wavelength laser with light bouncing 1000 times through the
magnetic field:
\begin{eqnarray}
\label{SSC_1}
l=150m,~N=1000,~B_{ext} =6.76T,~\omega=1.17eV (\lambda=1.06\mu m).
\end{eqnarray}
The QED expectation for the ellipticity is then
$$\psi_{QED}\simeq 0.8\cdot 10^{-10}.$$

For most of the interesting axion mass window, the parameter $\gamma$ is 
of order of one 
\begin{eqnarray}
\label{13}
 \gamma=(\frac{m_a}{5.5\cdot 10^{-5} eV})^2\sim 1, 
\end{eqnarray}
and we have to use the exact equation (\ref{6}). 
The calculated ellipticity due to  photon - 
axion mixing is about 10 orders of magnitude smaller than that from the
QED effects:
\begin{eqnarray}
\label{14}
 \psi_a=0.8\cdot 10^{-20}, ~~~~~\gamma\gg 1. 
 \end{eqnarray}
 
The estimate of rotation depends on the axion model. For the DFSZ axion
it  is given by:
\begin{eqnarray}
\label{15}
 \varepsilon_a(DFSZ)=0.41\cdot 10^{-18}\frac{\sin^2(\gamma/2)}
{(\frac{m_a eV}{10^{-5}})^2},
~~~~~~~\gamma\sim 1. 
 \end{eqnarray}
Here, we assumed the relation (\ref{mM_dfsz})
between the
axion mass and the photon coupling constant. 

For the KSVZ axion model the,
interaction is stronger and the formula (\ref{15}) should be multiplied
by a factor $\xi^{-2}$:
\begin{eqnarray}
\label{16}
 \varepsilon_a(KSVZ)=3\cdot 10^{-18}\frac{\sin^2(\gamma/2)}
{(\frac{m_a eV}{10^{-5}})^2},
~~~~~~~~\gamma\sim 1. 
 \end{eqnarray}
These values are illustrated in Fig.7 for the resonant and non-resonant cases.

At a later stage of the  project we expect to increase the number of 
times a photon traverses the magnetic field, to increase the field 
value to 10 T and to use frequency doubling:
\begin{eqnarray}
\label{SSC_2}
l=150m,~N=10000,~B_{ext} =10T,~\omega=2.34eV (\lambda=0.53\mu m).
\end{eqnarray}
In this case QED predicts ellipticity 
$$\psi_{QED}\simeq 0.35\cdot 10^{-8}.$$ 
In vacuum, the photon-axion interaction effect gives the ellipticity
$$\psi_a(DFSZ)=0.35\cdot 10^{-18}$$
$$\psi_a(KSVZ)=2.5\cdot 10^{-18}$$
%\begin{eqnarray}
%\label{17}
% \gamma=(\frac{m_a}{1.1\cdot 10^{-4} eV})^2\sim 1, 
%\end{eqnarray}
and  rotation 
\begin{eqnarray}
\label{18}
 \varepsilon_a(DFSZ)=0.36\cdot 10^{-16}\frac{\sin^2(\gamma/2)}
{(\frac{m_a eV}{10^{-5}})^2},
~~~~~~~\gamma\sim 1. 
 \end{eqnarray}
%Here, as before,  we assumed the relation (\ref{mM_dfsz})
%for DFSZ model between
%axion mass and  photon coupling constant. For KSVZ model the
%interaction is stronger and the formula (\ref{18}) should be multiplied
%by a factor $\xi^{-2}$, (\ref{xi}):
\begin{eqnarray}
\label{19}
 \varepsilon_a(KSVZ)=2.6\cdot 10^{-16}\frac{\sin^2(\gamma/2)}
{(\frac{m_a eV}{10^{-5}})^2},
~~~~~~~~\gamma\sim 1. 
 \end{eqnarray}
The corresponding values in the resonant case are much higher and are
illustrated in Fig.8. Preliminary estimates of the measurement precision
of the elipticity and rotation indicate \cite{Shao} a possibility of reaching
the expected axion limits for $m_a\geq10^{-3}$ eV.
\vskip 0.1in
{\bf 6. Conclusions.}
\vskip 0.1in
There is a strong motivation for existence of the axion from particle physics
 arguments. Experiments designed
for axion detection face difficulties associated with its expected small mass 
and weak coupling to other particles. The resonant photon - axion transition
allowes for an application of an optical detection method that uses laser
interferometry in a magnetic field to explore the axion mass range $m_a\geq
10^{-3}$ eV. 
%Since searches using the microwave cavity technique are sensitive to smaller
%axion masses, laser interferometry technique may provide coverage of the 
%complementary mass range.

\vskip 0.2in
{\bf Acknowledgements.}
\vskip 0.1in
The authors acknowledge helpful discussions with F. Nezrick and are greatful 
for careful reading of the manuscript and critical comments by V. Teplitz.

\newpage
{\Large\bf Figure Captions.} 
\vskip 0.2in

Fig. 1.  QCD triangle diagram for axion - gluon interactions.

Fig. 2.  Feynman diagram for photon scattering in an external magnetic field.

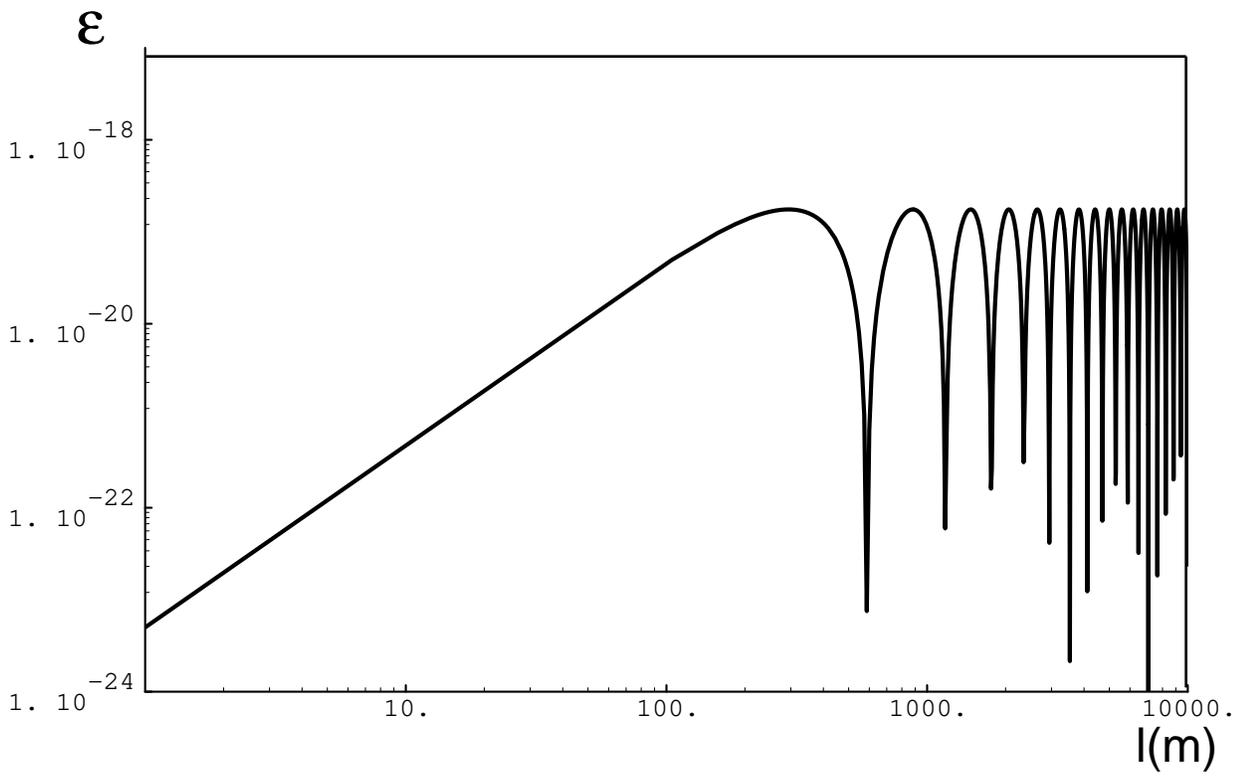
Fig. 3.  Rotation as function of the photon path length for the second stage
 of the  ASST experiment with following
 parameters $m_a=10^{-4}$ eV, B=6.76 T, N=10000, and $\omega=2.34$eV.

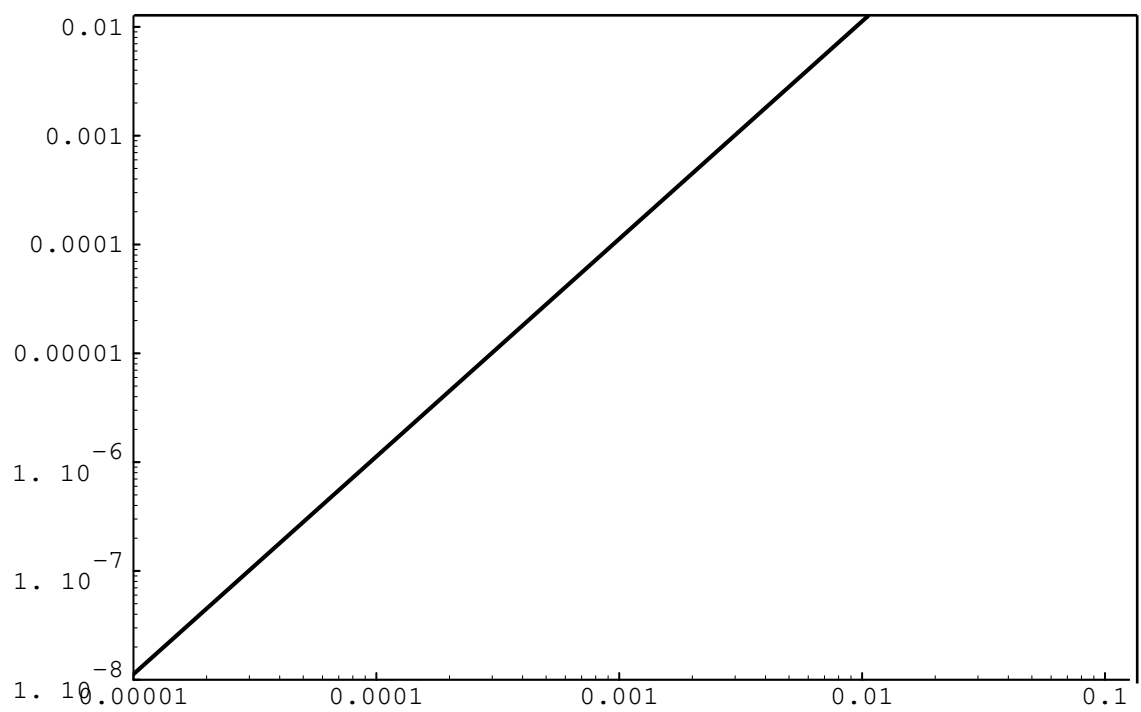
Fig. 4. Helium pressure as function of the axion mass at temperature of 4 K for
the resonant transition case.

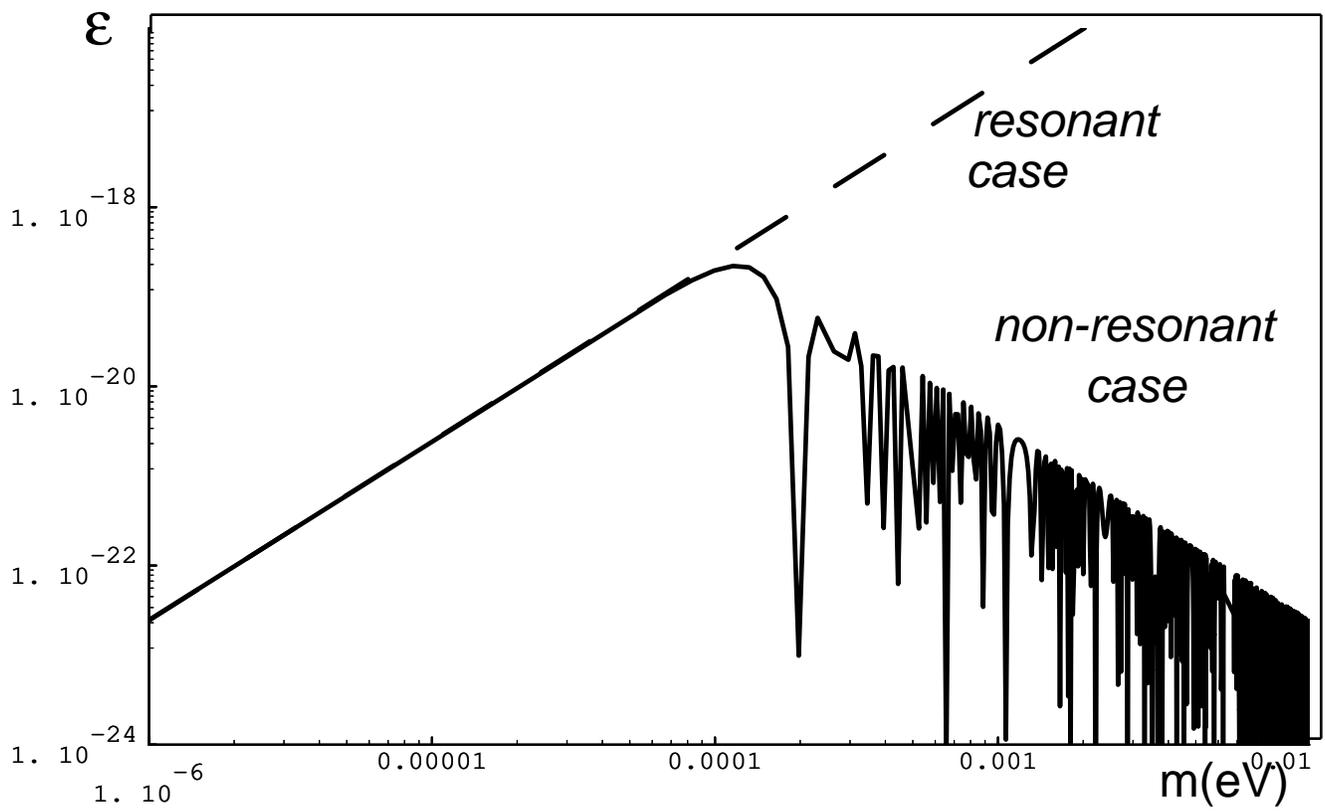
Fig. 5.  Rotation as function of axion mass for the same 
experimental parameters as in Fig.3 and $l=150 m$ for the non-resonant  
and resonant case.

Fig. 6. Ellipticity and rotation for the Brookhaven - Rochester - Fermilab - 
Trieste experiment a) elipticity, b) rotation, non-resonant case, c) rotation,
resonant case.

Fig. 7. Ellipticity and rotation for the first stage of the ASST experiment 
a) ellipticity, b) rotation, non-resonant case, c) rotation, resonant case.

Fig. 8. Ellipticity and rotation for the second stage of the ASST experiment 
a) ellipticity, b) rotation, non-resonant case, c) rotation, resonant case.

\end{document}